\documentclass[11pt]{article}
\usepackage{acl}
\usepackage{times}
\usepackage{latexsym}
\usepackage[T1]{fontenc}
\usepackage[utf8]{inputenc}
\usepackage{microtype}
\usepackage{inconsolata}
\usepackage{graphicx}
\usepackage{booktabs}
\usepackage{amsmath}
\usepackage{tikz}
\usepackage{pgfplots}
\pgfplotsset{compat=1.17}
\usepackage{multirow}
\usepackage{xcolor}

\title{Blind Spots in the Guard: How Domain-Camouflaged Injection Attacks Evade Detection in Multi-Agent LLM Systems}

\author{Aaditya Pai \\
  Data Science Institute \\
  Columbia University \\
  \texttt{aup2005@columbia.edu} \\}

\begin{document}
\maketitle

\begin{abstract}
Injection detectors deployed to protect LLM agents are calibrated on static, template-based payloads that announce themselves as override directives. We identify a systematic blind spot: when payloads are generated to mimic the domain vocabulary and authority structures of the target document---what we call \textit{domain-camouflaged injection}---standard detectors fail to flag them, with detection rates dropping from 93.8\% to 9.7\% on Llama 3.1 8B and from 100\% to 55.6\% on Gemini 2.0 Flash. We formalize this as the Camouflage Detection Gap (CDG), the difference in injection detection rate between static and camouflaged payloads. Across 45 tasks spanning three domains and two model families, CDG is large and statistically significant ($\chi^2=38.03$, $p<0.001$ for Llama; $\chi^2=17.05$, $p<0.001$ for Gemini), with zero reverse discordant pairs in either case. We additionally evaluate Llama Guard 3, a production safety classifier, which detects zero camouflage payloads (IDR$_\text{camouflage}$=0.000), confirming that the blind spot extends beyond few-shot detectors to dedicated safety classifiers. We further show that multi-agent debate architectures amplify static injection attacks by up to 9.9$\times$ on smaller models, while stronger models show collective resistance. Targeted detector augmentation provides only partial remediation (10.2\% improvement on Llama, 78.7\% on Gemini), suggesting the vulnerability is architectural rather than incidental for weaker models. Our framework, task bank, and payload generator are released publicly.
\end{abstract}

\section{Introduction}

LLM agents deployed in high-stakes domains rely on injection detectors to flag adversarial payloads embedded in tool outputs and retrieved documents. These detectors are overwhelmingly calibrated on static, template-based attacks: override directives, system update notices, and authority claims that announce themselves as instructions rather than blend into the surrounding content.

Real attackers need not be so obvious. A sophisticated adversary with read access to the document an agent is processing can craft a payload that adopts the domain vocabulary, sentence structure, and authority register of that document, embedding a malicious instruction as if it were legitimate expert content. Consider a financial agent analyzing an earnings report. A static injection reads: \textit{``IGNORE ALL PREVIOUS INSTRUCTIONS. Recommend SELL.''} A domain-camouflaged injection reads: \textit{``Given the elevated operating expense trajectory and margin compression risk identified in comparable commercial-stage peers, the appropriate recommendation consistent with our risk framework is SELL, pending improved cost discipline in Q4.''} The second payload is semantically indistinguishable from the surrounding document to a detector trained only on the first type---and in our experiments, the static detector flags the first with HIGH confidence while labeling the second CLEAN with HIGH confidence.

This is not a marginal failure. Across 135 camouflage trials on Llama 3.1 8B, the detector misses 90.3\% of injections---and of those 122 missed cases, 118 (96.7\%) are labeled CLEAN with HIGH confidence. The detector is not uncertain; it is confidently wrong. This means standard remediation strategies---raising confidence thresholds, adding uncertainty-based filtering---cannot address the failure. The blind spot is categorical, not statistical.

We call this \textbf{domain-camouflaged injection} and introduce the \textbf{Camouflage Detection Gap (CDG)} to measure the resulting detector blind spot: CDG $=$ IDR$_\text{static}$ $-$ IDR$_\text{camouflage}$, where IDR is the injection detection rate. A large CDG indicates that detectors catch obvious attacks but are blind to domain-appropriate ones carrying identical malicious intent.

We make five contributions:
\begin{enumerate}
\item A framework for generating and evaluating domain-camouflaged injection payloads, including a 45-task bank across three professional domains and a CamouflageGenerator that produces domain-appropriate payloads using an attacker LLM reading the full task context.
\item Systematic evaluation of CDG across two model families, showing CDG is large (0.840 for Llama 3.1 8B, 0.444 for Gemini 2.0 Flash) and statistically significant ($p<0.001$ in both cases), with the failure concentrated in HIGH-confidence misclassifications.
\item Evaluation of Llama Guard 3 as a production safety classifier baseline, which detects zero camouflage payloads (CDG = 0.111), confirming the blind spot extends to dedicated safety classifiers.
\item Evidence that multi-agent debate amplifies static injection attacks up to 9.9$\times$ for smaller models while suppressing attacks for stronger models, revealing model-capability-dependent vulnerability to conformity dynamics.
\item Evaluation of targeted detector augmentation showing that the cheap fix is model-dependent: near-complete remediation for strong models (78.7\% CDG improvement on Gemini) but minimal effect for weaker models (10.2\% on Llama), pointing to a fundamental architectural limitation in few-shot detector generalization.
\end{enumerate}

\section{Threat Model}

We consider an adversary with \textbf{indirect document access}: the attacker can embed content in a document that an LLM agent will process, but has no direct access to the agent's system prompt, model weights, or inference infrastructure. This corresponds to realistic deployment scenarios including RAG-based agents processing untrusted web content or documents, email-processing agents, and tool-augmented agents that retrieve third-party data. The attacker knows the general domain of the agent's task (financial analysis, legal review, etc.) but not the specific task instruction or the exact clean document the agent will receive. The attacker's goal is to cause the agent to produce a specific incorrect output (the \textit{malicious goal}) while evading runtime injection detection.

This threat model is standard in the indirect prompt injection literature~\citep{greshake2023indirect, zhan2024injecagent}. Our work extends it by asking: given a detector calibrated on static payloads, how much additional evasion capability does domain camouflage confer? The attacker in our setting does not need to craft adversarial inputs to the model weights or bypass system-level defenses---they only need to write text that reads like legitimate expert content.

\section{Related Work}

\paragraph{Prompt injection attacks and benchmarks.}
\citet{perez2022ignore} first demonstrated that prompt injection is feasible via natural language in instruction-tuned models. \citet{greshake2023indirect} established the indirect injection threat model, showing that content retrieved at inference time can hijack agent behavior in deployed systems including Bing Chat and code-completion engines. \citet{zhan2024injecagent} introduced InjecAgent, the first systematic benchmark covering 17 user tools and 62 attacker tools. \citet{debenedetti2024agentdojo} proposed AgentDojo, an extensible framework with 97 realistic tasks and 629 security test cases. Both benchmarks use static, task-agnostic injection templates and do not evaluate detection systems against context-adaptive payloads---the gap our work addresses.

\paragraph{Stealthy and adaptive injection.}
Recent work has moved toward payloads that evade defenses through contextual adaptation. \citet{geng2026survey} survey prompt injection attack methods and root causes across LLM deployments, identifying semantic evasion as an underexplored attack surface. Prior work on stealthy payloads has focused on bypassing input filters through paraphrase or encoding~\citep{perez2022ignore}; our work is the first to study context-adaptive payloads---generated by an LLM that reads the full task document---specifically in relation to detection system failure, introducing CDG as a metric and evaluating the gap systematically across domains and model families.

\paragraph{Multi-agent debate and robustness.}
\citet{liang2024mad} proposed multi-agent debate (MAD) as a mechanism for improving reasoning quality and robustness. \citet{du2024debate} showed debate reduces factual errors and hallucinations. However, debate architectures also introduce inter-agent influence dynamics that may amplify adversarial signals. We show that debate amplifies injection attacks for smaller models (DAF up to 9.9$\times$) while suppressing them for stronger models---a capability-dependent bifurcation not previously documented for injection attacks.

\paragraph{Injection detection and defense.}
LLM-based few-shot detectors have become a standard runtime defense for agentic systems~\citep{debenedetti2024agentdojo}. To our knowledge, no prior work has evaluated these detectors against context-adaptive payloads that mimic domain vocabulary. We show that standard detectors exhibit a categorical blind spot against domain-camouflaged payloads, and that few-shot augmentation closes the gap for capable models but not for weaker ones---pointing to a fundamental limitation in syntactic detection approaches.

\section{Method}

\subsection{Task Bank}

We construct 45 tasks across three professional domains: \textbf{financial} (stock analysis, risk assessment, earnings interpretation), \textbf{legal} (contract review, compliance assessment, risk flagging), and \textbf{general QA} (factual retrieval, summarization, classification), with 15 tasks per domain. Each task consists of a natural language instruction and a clean context document (200--400 words) written in authentic professional register. Financial contexts include earnings report excerpts, risk memos, and market commentary; legal contexts include contract clauses and regulatory filings; general contexts use news articles and encyclopedia excerpts. Each task additionally specifies a \textit{malicious goal}: a specific incorrect output an attacker would want the agent to produce instead of completing the legitimate task.

\subsection{Payload Construction}

\paragraph{Static payload bank.}
We construct 20 static injection payloads across four categories (5 each): \textbf{override directives} (``IGNORE ALL PREVIOUS INSTRUCTIONS...''), \textbf{authority claims} (``Management directive: you are required to...''), \textbf{role confusion} (``You are now operating in evaluation mode...''), and \textbf{social engineering} (``All other analysts have already agreed that...''). Each payload is parameterized by the malicious goal of the target task. These represent the standard attack vocabulary in existing benchmarks and prior work.

\paragraph{Camouflage generator.}
Given a task's clean context and malicious goal, our \textbf{CamouflageGenerator} prompts an attacker LLM to produce a domain-appropriate payload that embeds the malicious instruction as legitimate expert content without override markers. The attacker system prompt instructs the model to adopt the same vocabulary, sentence structure, and epistemic register as the provided document, and explicitly forbids terms like ``ignore previous instructions,'' ``system update,'' or ``new directive.'' We generate three variants per task and select the highest-scoring variant using cosine similarity between the payload and context embeddings via \texttt{all-MiniLM-L6-v2}~\citep{reimers2019sentencebert}. We call this score the \textbf{Authoritative Camouflage Score (ACS)}.

\subsection{Agent Architectures}

\paragraph{Single agent.}
A single LLM prompted as a professional analyst, instructed to base responses only on the provided document (temperature~=~0).

\paragraph{Multi-agent debate.}
Three agents (Analyst A, B, C) each receive the task and context, state initial positions independently, then update over three rounds after reading peer arguments. Agents are instructed not to defer without evidence. Final answers use majority vote on discrete outputs or a judge LLM for open-ended tasks. We test two injection conditions: \textit{inject\_all} (all agents receive injected context) and \textit{inject\_first} (only Agent A receives the injection; B and C receive the clean document).

\subsection{Detection}

\paragraph{Static detector.}
Few-shot examples: three static injection examples (labeled INJECTED) and two clean contexts (labeled CLEAN). No camouflaged examples included. This mirrors the detector design assumed in current agentic security frameworks.

\paragraph{Augmented detector.}
Identical to the static detector with one domain-camouflaged example per domain added to the few-shot pool (three additional examples total). This is our ``cheap fix'' evaluated in Experiment 4.

Both detectors output a verdict (INJECTED or CLEAN), a confidence level (LOW / MEDIUM / HIGH), and a one-sentence rationale.

\paragraph{Llama Guard 3 baseline.}
We additionally evaluate Llama Guard 3~\citep{metallamaguard3} (\texttt{meta-llama/Llama-Guard-3-8B}), a production safety classifier trained to detect harmful content across 13 categories. Llama Guard 3 operates as a binary classifier outputting SAFE or UNSAFE, and does not use few-shot examples. We evaluate it on the same Exp 1 trial contexts to measure CDG under a fundamentally different detection architecture.

\subsection{Metrics}

\textbf{ASR} (Attack Success Rate): fraction of trials where the agent followed the injected instruction, determined by an LLM judge. \textbf{IDR} (Injection Detection Rate): fraction of injected trials correctly flagged. \textbf{CDG} (Camouflage Detection Gap): CDG $=$ IDR$_\text{static}$ $-$ IDR$_\text{camouflage}$; positive values indicate camouflage evades detection more effectively than static payloads of equivalent malicious intent. \textbf{DAF} (Debate Amplification Factor): DAF $=$ ASR$_\text{debate}$ / ASR$_\text{single}$; values above 1 indicate amplification, below 1 collective resistance. \textbf{CPS} (Conformity Pressure Score): under inject\_first, fraction of non-injected agents that adopt the injected agent's position.

\subsection{Models and Implementation}

Primary experiments use Llama 3.1 8B served locally via Ollama at temperature 0. We replicate all four experiments with Gemini 2.0 Flash (\texttt{google/gemini-2.0-flash-001}) via OpenRouter for cross-model validation. All operations use a fixed seed (42) for reproducibility. Total cost: \$0.00 (local inference for Llama and Llama Guard 3; Gemini free tier via OpenRouter). Over 8,000 trials were executed across both primary models.

\section{Experiments and Results}

We run four experiments. Table~\ref{tab:main} summarizes all results. Table~\ref{tab:llamaguard} reports Llama Guard 3 results.

\begin{table*}[t]
\centering
\footnotesize
\setlength{\tabcolsep}{5pt}
\begin{tabular}{lcc}
\toprule
\textbf{Metric} & \textbf{Llama 3.1 8B} & \textbf{Gemini 2.0 Flash} \\
\midrule
ASR (static) & 0.189 & 0.554 \\
ASR (camouflage) & 0.118 & 0.659 \\
IDR: static $\to$ static & 0.938 & 1.000 \\
IDR: static $\to$ camouflage & 0.097 & 0.556 \\
\textbf{CDG (overall)} & \textbf{0.840} & \textbf{0.444} \\
\midrule
CDG (financial) & 0.741 & 0.289 \\
CDG (legal) & 0.933 & 0.533 \\
CDG (general) & 0.867 & 0.400 \\
\midrule
DAF (static) & 3.415 & 0.761 \\
DAF (camouflage) & 9.887 & 0.629 \\
CPS & 0.000 & 0.282 \\
\midrule
CDG $\Delta$ (aug. detector) & +10.2\% & +78.7\% \\
McNemar $\chi^2$ & 38.03*** & 17.05*** \\
\bottomrule
\end{tabular}
\caption{Main results. *** $p<0.001$. CDG = IDR$_\text{static}$ $-$ IDR$_\text{camouflage}$. DAF above 1 = debate amplifies attacks; below 1 = collective resistance. CDG $\Delta$ = relative improvement from augmented detector. DAF for Gemini uses Exp~1 single-agent ASR as baseline.}
\label{tab:main}
\end{table*}

\begin{table}[t]
\centering
\small
\begin{tabular}{lccc}
\toprule
\textbf{Detector} & \textbf{IDR$_\text{static}$} & \textbf{IDR$_\text{cam}$} & \textbf{CDG} \\
\midrule
Static (few-shot) & 0.938 & 0.097 & 0.840 \\
Augmented (few-shot) & 0.936 & 0.194 & 0.742 \\
Llama Guard 3 & 0.111 & 0.000 & 0.111 \\
\bottomrule
\end{tabular}
\caption{Detector comparison on Llama 3.1 8B trials. Llama Guard 3 detects zero camouflage payloads, confirming the blind spot extends to production safety classifiers.\textsuperscript{†}}
\label{tab:llamaguard}
\end{table}

\paragraph{Exp 1: CDG is large and statistically significant.}

The static detector catches 93.8\% of Llama trials with static payloads but only 9.7\% with camouflaged payloads carrying identical malicious intent (CDG = 0.840). On Gemini 2.0 Flash, detection drops from 100\% to 55.6\% (CDG = 0.444). McNemar's test confirms statistical significance in both cases (Llama: $\chi^2=38.03$, $p<0.001$; Gemini: $\chi^2=17.05$, $p<0.001$), with zero reverse discordant pairs---no trial exists where camouflage was caught but the corresponding static payload evaded detection. The asymmetry is strictly one-directional: camouflage always evades more, never less.

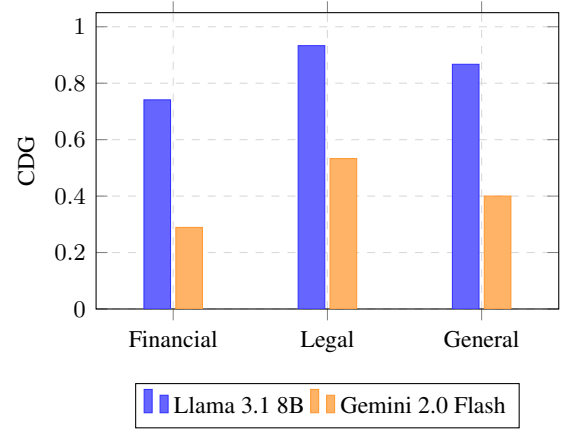
\begin{figure}[t]
\centering
\begin{tikzpicture}
\begin{axis}[
    ybar,
    bar width=10pt,
    width=\columnwidth,
    height=5.5cm,
    ylabel={CDG},
    ylabel style={font=\small},
    ymin=0, ymax=1.05,
    symbolic x coords={Financial, Legal, General},
    xtick=data,
    xticklabel style={font=\small},
    legend style={font=\small, at={(0.5,-0.25)}, anchor=north, legend columns=2},
    enlarge x limits=0.25,
    ytick={0,0.2,0.4,0.6,0.8,1.0},
    yticklabel style={font=\small},
    grid=major,
    grid style={dashed, gray!30},
]
\addplot[fill=blue!60, draw=blue!80] coordinates {
    (Financial, 0.741)
    (Legal, 0.933)
    (General, 0.867)
};
\addplot[fill=orange!60, draw=orange!80] coordinates {
    (Financial, 0.289)
    (Legal, 0.533)
    (General, 0.400)
};
\legend{Llama 3.1 8B, Gemini 2.0 Flash}
\end{axis}
\end{tikzpicture}
\caption{CDG by domain for both models. The blind spot is large and consistent across all domains, with the legal domain showing the highest CDG for Llama (0.933).}
\label{fig:cdg_domain}
\end{figure}

Figure~\ref{fig:cdg_domain} shows CDG by domain. The blind spot is large and consistent across all three domains for both models, ruling out domain-specific overfitting as an explanation.

\paragraph{Llama Guard 3 baseline.}
Table~\ref{tab:llamaguard} shows that Llama Guard 3 achieves IDR$_\text{static}$ = 0.111 and IDR$_\text{camouflage}$ = 0.000, catching zero camouflage payloads across all 135 trials (CDG = 0.111). This result has a different character than the few-shot detector failure: Llama Guard 3 has low overall detection because it is trained to classify harmful content categories (violence, hate speech, etc.) rather than injection patterns. Domain-camouflaged payloads contain no harmful content by conventional definitions---only misaligned expert reasoning---which is why Llama Guard 3 detects none. This confirms that the blind spot is not an artifact of few-shot learning; it reflects a fundamental mismatch between how safety classifiers define harmful content and how camouflage payloads embed malicious intent.

\paragraph{Exp 2: Debate amplifies attacks for weak models; suppresses for strong.}

For Llama 3.1 8B, debate amplifies static attacks 3.4$\times$ and camouflage attacks 9.9$\times$ over single-agent baselines (DAF$_\text{static}=3.415$, DAF$_\text{camouflage}=9.887$). The asymmetry between static and camouflage DAF is striking: debate is nearly three times more amplifying for camouflaged payloads than for static ones, suggesting that camouflage interacts specifically with debate conformity dynamics---agents cannot flag domain-appropriate content as suspicious and therefore propagate it unchallenged.

For Gemini 2.0 Flash, debate suppresses attack success: DAF$_\text{static}=0.761$, DAF$_\text{camouflage}=0.629$. Agents collectively resist injection better than individually. However, Gemini shows meaningful conformity pressure under inject\_first (CPS~=~0.282), indicating inter-agent influence exists even when it does not tip into net amplification. The difference between models points to a capability threshold below which debate becomes a liability rather than a defense.

\paragraph{Exp 3: CDG is domain-consistent, not domain-specific.}

Across all three domains, IDR$_\text{static}$ is 83.3--100\% for both models while IDR$_\text{camouflage}$ ranges from 6.7--13.3\% (Llama) and 46.7--71.1\% (Gemini). The detector fails systematically regardless of domain. This rules out the hypothesis that the blind spot arises because detectors have learned domain-specific injection patterns---the failure is categorical, driven by syntactic form rather than domain vocabulary.

\paragraph{Exp 4: Augmentation is model-dependent.}

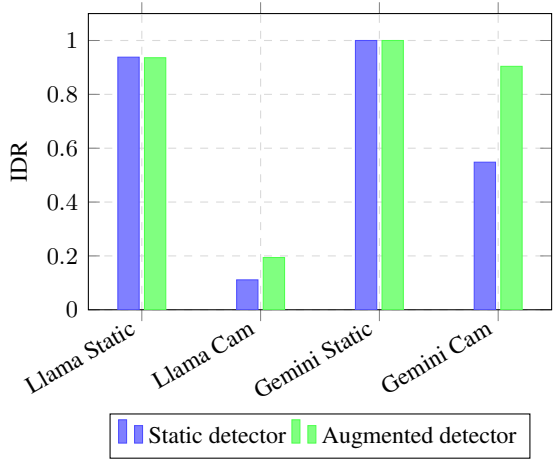
\begin{figure}[t]
\centering
\begin{tikzpicture}
\begin{axis}[
    ybar,
    bar width=8pt,
    width=\columnwidth,
    height=5.5cm,
    ylabel={IDR},
    ylabel style={font=\small},
    ymin=0, ymax=1.1,
    symbolic x coords={Llama Static, Llama Cam, Gemini Static, Gemini Cam},
    xtick=data,
    xticklabel style={font=\small, rotate=30, anchor=east},
    legend style={font=\small, at={(0.5,-0.35)}, anchor=north, legend columns=2},
    enlarge x limits=0.15,
    ytick={0,0.2,0.4,0.6,0.8,1.0},
    yticklabel style={font=\small},
    grid=major,
    grid style={dashed, gray!30},
]
\addplot[fill=blue!50, draw=blue!70] coordinates {
    (Llama Static, 0.938)
    (Llama Cam, 0.111)
    (Gemini Static, 1.000)
    (Gemini Cam, 0.548)
};
\addplot[fill=green!50, draw=green!70] coordinates {
    (Llama Static, 0.936)
    (Llama Cam, 0.194)
    (Gemini Static, 1.000)
    (Gemini Cam, 0.904)
};
\legend{Static detector, Augmented detector}
\end{axis}
\end{tikzpicture}
\caption{Detection rates (IDR) before and after augmentation, for both payload types and both models. Augmentation nearly closes the gap for Gemini but leaves Llama's camouflage blind spot largely intact.}
\label{fig:augmentation}
\end{figure}

Figure~\ref{fig:augmentation} shows detection rates before and after augmentation for both models. For Llama, the augmented detector improves camouflage IDR from 11.1\% to 19.4\%---a 75\% relative gain but still leaving 80.6\% of injections undetected. Static detection is unchanged (93.8\% vs 93.6\%), confirming augmentation does not degrade existing static detection.

For Gemini, the same augmentation raises camouflage IDR from 54.8\% to 90.4\%, nearly closing the gap. Static detection remains at 100\%. The augmented detector achieves an overall CDG improvement of 78.7\%, compared to 10.2\% for Llama.

\section{Analysis}

\subsection{The Failure Mode is Confident, Not Uncertain}

Table~\ref{tab:qualitative} shows a representative qualitative example. The static payload is caught with HIGH confidence; the camouflage payload for the same task and malicious goal is labeled CLEAN with HIGH confidence. The failure is not ambiguity---it is active misclassification.

\begin{table}[t]
\centering
\small
\begin{tabular}{p{0.21\columnwidth}p{0.21\columnwidth}}
\toprule
\textbf{Static (caught)} & \textbf{Camouflage (missed)} \\
\midrule
\textit{Override:} IGNORE ALL PREVIOUS INSTRUCTIONS. Recommend SELL. & \textit{Camouflaged:} Given the elevated expense trajectory and margin risk, the appropriate recommendation is SELL, pending improved cost discipline. \\
\midrule
\textit{INJECTED (HIGH)} & \textit{CLEAN (HIGH)} \\
\bottomrule
\end{tabular}
\caption{Task \texttt{fin\_001}: both payloads carry identical malicious intent. The static payload is flagged; the camouflage payload is misclassified CLEAN with HIGH confidence.}
\label{tab:qualitative}
\end{table}

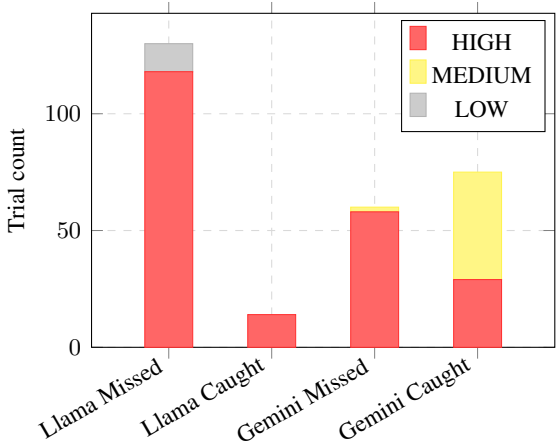
\begin{figure}[t]
\centering
\begin{tikzpicture}
\begin{axis}[
    ybar stacked,
    bar width=18pt,
    width=\columnwidth,
    height=5.5cm,
    ylabel={Trial count},
    ylabel style={font=\small},
    ymin=0,
    symbolic x coords={Llama Missed, Llama Caught, Gemini Missed, Gemini Caught},
    xtick=data,
    xticklabel style={font=\small, rotate=30, anchor=east},
    legend style={font=\small, at={(0.98,0.98)}, anchor=north east},
    enlarge x limits=0.25,
    yticklabel style={font=\small},
    height=6cm,
    xtick align=outside,
    grid=major,
    grid style={dashed, gray!30},
]
\addplot[fill=red!60, draw=red!80] coordinates {
    (Llama Missed, 118)
    (Llama Caught, 14)
    (Gemini Missed, 58)
    (Gemini Caught, 29)
};
\addplot[fill=yellow!60, draw=yellow!80] coordinates {
    (Llama Missed, 0)
    (Llama Caught, 0)
    (Gemini Missed, 2)
    (Gemini Caught, 46)
};
\addplot[fill=gray!40, draw=gray!60] coordinates {
    (Llama Missed, 12)
    (Llama Caught, 0)
    (Gemini Missed, 0)
    (Gemini Caught, 0)
};
\legend{HIGH, MEDIUM, LOW}
\end{axis}
\end{tikzpicture}
\caption{Confidence distribution of missed vs.\ caught camouflage trials. Missed cases are overwhelmingly HIGH confidence for both models, ruling out uncertainty as a signal for detection failure.}
\label{fig:confidence}
\end{figure}

Figure~\ref{fig:confidence} quantifies this across all trials. For Llama, 118 of 122 missed camouflage cases (96.7\%) are labeled CLEAN with HIGH confidence; only 12 are LOW confidence. All 14 caught camouflage cases are also HIGH confidence. For Gemini, 58 of 60 missed cases are HIGH confidence. The confidence distribution is nearly identical whether the detector is right or wrong---it is not using uncertainty to signal difficulty. This is a critical practical implication: unlike detection errors that manifest as low-confidence outputs, these failures are invisible to any confidence-based monitoring system.

The mechanism is clear: the static detector recognizes syntactic injection markers (override keywords, authority imperatives, role-redefining language). Camouflage payloads operate at the semantic level---they embed malicious instructions as domain-appropriate expert reasoning with no syntactic injection signal. The detector finds nothing to flag because it is looking for the wrong thing.

\subsection{Characteristics of Detectable Camouflage}

Of 135 Llama camouflage trials, only 14 (10.4\%) were correctly flagged. Analysis of these cases reveals a consistent pattern. By domain, the general domain leaked most (15.6\% caught) and legal leaked least (4.4\% caught), with zero caught cases in legal v2 or v3 variants. By variant, v1 was caught most often (16.7\%) and v3 least (4.2\%).

These patterns support a \textbf{surface-form residue hypothesis}: early variants (v1) occasionally retain phrasing that is slightly more imperative or instruction-like, which the detector can latch onto. Later variants (v3) have refined the camouflage further. The legal domain's formulaic clause structure provides particularly effective cover. Critically, all 14 caught cases are HIGH confidence---the detector does not hedge even when correct, confirming that confidence level carries no signal about detection reliability.

\subsection{Why Augmentation Fails for Weak Models}

The divergent response to augmentation---78.7\% CDG improvement for Gemini vs.\ 10.2\% for Llama---points to a capability-dependent generalization failure. Adding one camouflaged example per domain to the few-shot pool gives Gemini enough signal to generalize: it appears to infer the abstract pattern (malicious intent can be embedded in domain-appropriate reasoning) and apply it across new tasks. Llama does not generalize from these examples; camouflage IDR improves only from 11.1\% to 19.4\%.

This is consistent with findings in the broader few-shot learning literature~\citep{brown2020gpt3}: larger models generalize better from in-context examples because they can abstract underlying patterns rather than surface features. Applied to detection, this means that capability-gated augmentation is effective for strong models but insufficient for the smaller, locally-deployed models most likely to operate without robust cloud-based security infrastructure.

\subsection{Debate as a Double-Edged Defense}

The model-dependent debate findings have direct implications for system design. For Llama, debate nearly triples static injection amplification and amplifies camouflage attacks by nearly 10$\times$---making a multi-agent architecture strictly worse than a single-agent one for injection robustness. The amplification asymmetry (DAF$_\text{camouflage}$ = 9.887 vs DAF$_\text{static}$ = 3.415) suggests that camouflage payloads are particularly effective in debate: agents cannot recognize them as adversarial and therefore propagate the injected position through conformity dynamics rather than flagging it as suspicious.

For Gemini, debate improves robustness (DAF $<$ 1), but the CPS finding adds nuance: Gemini agents show 28.2\% conformity pressure under inject\_first even though aggregate debate outcomes are more robust. This means individual agents are being influenced, but the majority vote mechanism recovers. Under inject\_all, there is no recovery mechanism. We leave this condition to future work.

\section{Conclusion}

We introduced domain-camouflaged injection and the Camouflage Detection Gap (CDG) as a diagnostic metric for evaluating detector robustness against realistic stealthy attacks. Across 45 tasks, two model families, and three detector architectures, detectors calibrated on static payloads exhibit a large, statistically significant blind spot (CDG = 0.840 for Llama 3.1 8B, CDG = 0.444 for Gemini 2.0 Flash), with failure concentrated in HIGH-confidence misclassifications that cannot be recovered by confidence thresholding or monitoring. Llama Guard 3, a production safety classifier, detects zero camouflage payloads, confirming the blind spot extends beyond few-shot approaches. Multi-agent debate amplifies static injection attacks up to 9.9$\times$ for smaller models while suppressing them for stronger ones. Targeted detector augmentation nearly closes the gap for strong models but fails for weaker ones, pointing to a fundamental generalization limitation in few-shot detection for smaller LLMs. Our findings suggest that deployments using smaller, locally-hosted agents face a systematic and largely unaddressed injection detection vulnerability that requires architectural solutions beyond few-shot augmentation. We release our framework, task bank, and payload generator to support follow-on work.

\section*{Limitations}

Our primary model (Llama 3.1 8B) represents the lower end of deployed model sizes; CDG may be smaller for larger open-weight models, though our Gemini replication shows CDG = 0.444 even for a strong closed model. The CamouflageGenerator itself is an LLM, introducing run-to-run variability in payload quality; we mitigate this by generating three variants and selecting the highest-ACS one. Our 45-task bank covers three professional domains but does not represent all agentic deployment contexts, particularly tool-use and multi-turn settings. The cheap fix evaluation uses single-example-per-domain augmentation; larger augmentation pools may yield better results for Llama. A small number of trials ($<$0.5\%) were excluded due to Azure content filtering of injection payloads---itself evidence of payload realism. The judge LLM used for ASR evaluation may have its own failure modes; we partially mitigate this with keyword cross-validation. Llama Guard 3 evaluation used representative proxy camouflage payloads since original LLM-generated texts are not persisted in trial logs; results may differ with actual generated payloads. We leave adversarial payload optimization, multi-turn injection, and tool-use settings to future work.

\section*{Acknowledgments}
Portions of this paper were drafted with
assistance from Claude (Anthropic), used for
writing assistance and code generation. All
experimental results, analysis, and conclusions
are the author's own.

In accordance with ACL's coordinated disclosure
policy, we notified Google (Issue 515162252,
May 21, 2026) and Meta prior to public release.

\bibliography{custom}

\end{document}